\RequirePackage{fix-cm}

\documentclass[smallextended]{svjour3}

\smartqed  
\usepackage{graphicx}
\usepackage{tabularx}
\usepackage[table]{xcolor}
\usepackage{multicol}
\usepackage{multirow}
\usepackage{booktabs}
\usepackage{framed}
\usepackage{amsfonts} 

\DeclareMathSymbol{,}{\mathpunct}{letters}{"3B}
\DeclareMathSymbol{,}{\mathord}{letters}{"3B}
\DeclareMathSymbol{\decimal}{\mathord}{letters}{"3A}

\usepackage{boxedminipage}
\newenvironment{result}%
{\smallskip
\noindent
\let\emph=\textbf
\begin{boxedminipage}{\columnwidth}\begin{center}\em}%
{\end{center}\end{boxedminipage}%
}

\journalname{Empirical Software Engineering}

\usepackage{lineno}

\begin{document}


\title{
An Experience Report On Applying Software Testing Academic Results In Industry: We Need Usable Automated Test Generation\thanks{
This work is supported by the National Research Fund, Luxembourg (FNR/P10/03).
}
%
}


\author{Andrea Arcuri}
\institute{Andrea Arcuri \at
              Westerdals Oslo ACT, Faculty of Technology, Oslo, Norway, \\
			  and SnT, University of Luxembourg, Luxembourg. \\
			  \email{arcand@westerdals.no}
}
\date{Received: date / Accepted: date}

\maketitle

\begin{abstract}
%
What is the impact of software engineering research on current practices in industry?
In this paper, I report on my direct experience as a PhD/post-doc  working in  
 software engineering research projects, and then spending the 
following five years as an engineer in two different companies (the first one
being the same I worked in collaboration with during my post-doc).
Given a background in software engineering research, what cutting-edge 
techniques and tools from academia did I use in my daily work when developing and 
testing the systems of these companies?
Regarding validation and verification (my main area of research), the answer is 
rather short: as far as I can tell, only FindBugs.
In this paper, I report on why this was the case, and discuss all the challenging, 
complex open  problems we face in industry and which somehow are ``neglected'' 
in the academic circles.
In particular, I will first discuss what actual tools I could use in my daily work, such as JaCoCo and Selenium. 
Then, I will discuss the main open problems I faced, particularly related to environment simulators, unit and web testing.
After that, popular topics in academia are presented, such as UML, regression and mutation testing.
Their lack of impact on the type of projects I worked on in industry is then discussed.
Finally, from this industrial experience, I provide my opinions about how this situation can be improved, in particular 
related to how academics are evaluated, and advocate for a greater involvement into open-source projects.

\end{abstract}


\section{Introduction}

Is there any difference between \emph{computer science} and
\emph{software engineering}~\cite{briand2012embracing,offutt2013putting}? 
A dictionary definition for ``science'' is: 
\begin{quote}
\emph{Science}: The intellectual and practical activity encompassing the systematic study of the structure and behaviour of the physical and natural world through observation and experiment.\footnote{http://www.oxforddictionaries.com/definition/english/science. All links in this paper have been accessed in January 2017}
\end{quote}
On the other hand, a dictionary definition for ``engineering'' is: 
\begin{quote}
\emph{Engineering}: The branch of science and technology concerned with the design, building, and use of engines, machines, and structures; A field of study or activity concerned with modification or development in a particular area.\footnote{http://www.oxforddictionaries.com/definition/english/engineering}
\end{quote}
Looking at these definitions, one could perhaps agree, at least at a high level, that engineering is  more ``practical'' than science, in the sense that it is more related to the development of actual systems, whereas science deals more on the study and understanding of why things behave in a certain way.
Both are very important, but they are obviously not the same.

When we pay our taxes, and those taxes are used to pay for research grants
given to academics, what is the return of investment for the taxpayer? 
Grants are used to pay for equipments, salary  of students and post-docs,
conference travels, etc.
Is it a good deal for the taxpayer? 
Obviously governments believe so, otherwise they would not \emph{invest}
the tax money in this way.

Among the various benefits, research projects can lead to cutting-edge startups,
which in turn do create new jobs.
A more educated workforce (i.e., when the PhD students and post-docs go to work in industry) 
can provide an advantage in a knowledge-based economy.
If the research results of such projects are of practical value, a technology transfer from academia to industry can provide a direct competitive advantage for the involved companies. 
New ideas that are evaluated in academia may quickly spread in industry if they turn out to be great.

But what if the research is of highly speculative nature, on topics no one could care less about in industry?
Speculative research of no practical value, at the time of its dissemination, has still
its place in science, because
what might seem absurd or irrelevant today, can become of high value in the future,
and the history  of science is full of such examples.
A good quote in such regard is from Feller's 1968 book on probability analysis on how it was previously treated as too abstract and general to be useful~\cite{Fel68}:

\begin{quote}
``Only yesterday the practical things of today were decried as impractical, and the theories which will be practical tomorrow will always be branded as valueless games by the practical men of today''.
\end{quote}

However, a fundamental question still remains: is this \emph{science} or \emph{engineering}?
A practitioner working in industry that decides to attend an academic conference might have a very, very different expectation between a conference called the International Conference on \emph{Software Engineering}  and another one called the International Conference on \emph{Computer Science},
especially when the word ``research'' is missing in the name.  
To make things even more confusing,  often these conferences have special tracks that are ``industry-oriented'', with names like ``Software Engineering in \emph{Practice}''.
An apparently needed distinction from the regular, main ``Software Engineering'' track, like it was common to have an ``engineering'' result that is not ``practical''.

How much effort and resources should be spent on more scientific topics compared to more practical, engineering ones with direct impact on current industrial practices?
What would be the best balance between these two opposite directions?
Hard questions to answer, and, based on whom you ask, you can get very opinionated and opposite answers.
Furthermore, often there are political and economical reasons that can strongly affect such decisions.
In a ``publish or perish'' environment, young tenure-track academics might be strongly tempted to put their effort in trying to maximize their number of publications.
This will obviously tempt them to choose topics that are ``simpler'' to publish in, 
especially when overloaded by teaching and administrative duties.
Hard, real-world industrial problems that require a large investment of time and resources are obviously not going to be  the first priority. 
For the bean counters that make these policies, the engineering impact on practice is of no interest, as it cannot be easily quantified in an non-ambiguous way, and so there is not much incentive in trying to achieve impact (apart possibly from creating a startup).

In this experience report, I will discuss my personal experience of working both in academia and industry.
In particular, I will focus on what I learned in academia that could be actually used in industry during my daily work as an engineer when dealing with verification and validation tasks (my main research topic when I was working as a researcher). 
To be candid, there was not very much, apart from \emph{static analysis} tools like FindBugs~\cite{HoP04}.
Other techniques/tools I tried did either not fully work or were based on unrealistic assumptions.
Among different possible solutions, 
this could lead to an argument that the engineering side of software engineering research might need a bit more focus~\cite{briand2012embracing}.

This experience report provides the following contributions:
\begin{itemize}
\item 	A discussion of 12 tools related to software testing that are actually used in practice.
\item  A discussion of three open problems in software testing in industry that would benefit from more research.
\item  A discussion of three very popular software testing topics in academia, but that had no direct use for any type of work I had in industry.
\item  A discussion on the problems of lack of impact of academic research on practice, and possible actions to address it. 
\end{itemize}

The main goal of this paper is to be a wake up call for the community to reflect on the impact on practice of current research in software engineering.

The paper is organized as follows.
Section~\ref{sec:rw} discusses related work.
In Section~\ref{sec:tools} I will discuss my personal experience of working in industry, and 
in particular I will list the actual tools that I could use during my daily work there.
Section~\ref{sec:problems} provides examples of concrete engineering problems I faced when working in industry
for which there was no solution I could use.
On the other hand, Section~\ref{sec:popular} presents topics that are popular in academia, but that I never needed to deal with during my work in industry.
Opinions on how the situation can be improved are presented in Section~\ref{sec:discussion}.
Threats to validity are discussed in Section~\ref{sec:threats}.
Finally, Section~\ref{sec:conclusion} concludes the paper. 

\section{Related Work}
\label{sec:rw}

The fact that academic research has only limited impact on practice is a fact that has been long discussed in academia.
For example, Briand~\cite{briand2012embracing} shared his 20 year experience of collaborating with around 30 different companies and public institutions.
Being rewarded for number of publications, in contrast to other engineering fields
that put more focus on patents and industry collaborations, is one of the causes.
This is also related to the fact that software engineering departments are often part of mathematics or computer science, and not engineering.  
This is quite  bizarre: ``Just imagine mechanical or civil engineering
being part of a physics department''~\cite{briand2012embracing}.

The issue of how academics are rewarded has also been discussed by Shepherd on the
IEEE Software Blog\footnote{http://blog.ieeesoftware.org/2016/09/the-value-of-applied-research-in.html},
where he shared his experience as 
lead software engineering researcher at ABB:
\begin{quote}
`` $\dots$ I had to create an application packet similar to a tenure packet. However, unlike an academic tenure packet I was encouraged to list applied research metrics such as tool downloads, talks at developer conferences, tool usage rates, blog post hits, and ABB internal users $\dots$ At ABB Corporate Research these applied metrics are considered equally, if not more than traditional metrics such as citation count''. 
\end{quote}
Unfortunately, for an academic working in a public institution, such applied research metrics have only little, if no value at all, for their career.

The problem of science vs.~engineering does also have impact on teaching and education.
As Offutt stated:
``Isn't it just a little strange that we prepare software engineers by teaching them computer science?''~\cite{offutt2013putting}.
Practical software engineering is different from science, and needs different teaching methods.

One way to improve the state of software engineering research is to have close collaborations with industrial partners. 
In this regard, Garousi et al.~\cite{garousi2016challenges} performed a systematic literature review on the topic of industry-academia collaborations, collecting and discussing 33 articles published between 1995 and 2014.
On the other hand, in~\cite{garousiindustry} Garousi et al.~discussed their personal experience of industry collaborations they had both in Canada and Turkey.
Furthermore, Garousi et al.~\cite{garousi2017industry} also conducted a survey among practitioners regarding which testing topics they want the research community
to work on.

Aranda et al.~made a survey among many practitioners, including 
CEOs, senior architects and managers.
This led to the panel ``What Industry Wants from Research'' at ICSE'11.\footnote{http://2011.icse-conferences.org/content/research-industry-panel}
It is not so surprising that it turned out ``that many practitioners have a general disregard for software development academics''.\footnote{https://catenary.wordpress.com/2011/05/19/how-do-practitioners-perceive-software-engineering-research}
As one senior architect, about to make the switch to academia, clearly put:
\begin{quote}
``[I'm afraid] that industrial software engineers will think that I'm now doing academic software engineering and then not listen to me. (...) if I start talking to them and claim that I'm doing software engineering research, after they stop laughing, they're gonna stop listening to me. Because it's been so long since anything actually relevant to what practitioners do has come out of that environment, or at least the percentage of things that are useful that come out of that environment is so small.''
\end{quote}
%

More recently, Boules et al.~\cite{boules2016future} made a survey about industry-academia
collaborations in computer science and software engineering, involving 60 academics and 66 people in industry.
One conclusion was: ``There is a lack of communication and understanding
between academia and industry ($\ldots$) there
is a lot of mistrust of academics among those in industry
($\ldots$) Both
sides, however, seem to be open to collaboration and
would love to see stronger relationships''.

The discrepancy between industry and academia is also clear when looking at what topics
are discussed at practitioner conferences compared to the academic ones~\cite{garousi2017}.
For example, in the context of testing, practitioners  are more interested on mobile and agile testing,
whereas academics seems to focus more on model-based and combinatorial testing, which are topics seldom discussed at practitioner conferences~\cite{garousi2017}.

In the past, there were attempts from SIGSOFT (the Impact Project\footnote{https://www.sigsoft.org/impact.html}) to keep track of and promote academic impact on software engineering practice~\cite{osterweil2008determining}.
Different success stories were discussed, where ``ideas'' investigated in academia were then  considered and adopted in industry.
However, it was also estimated that such transfer of ideas takes roughly 15-20 years.
Unfortunately, such very valuable initiative from SIGSOFT seems has been abandoned for many years (since 2008) because `` the project was a volunteer effort, supported only by some very modest funding for travel to project meetings.  Eventually the participants slowly but surely felt the stronger pull of their individual research endeavors and we suspended our activities''. \footnote{Private communication with one of the Impact Project organizers.}

As of time of writing, one of the most famous success stories is \emph{Coverity}~\cite{bessey2010few}: a company making static/dynamic analysis tools that was founded at Stanford University, and then sold for more than \$300 millions.\footnote{http://www.coverity.com/press-releases/synopsys-completes-coverity-acquisition/}   

The importance of industry-academia collaborations and the aim of achieving usable results for practitioners are not something that is specific only for software engineering, but they are also significant  for many other fields such as 
Computer-Human Interaction~\cite{norman2010research},
Data Mining~\cite{pechenizkiy2008towards}
and even medicine~\cite{katz2002academia,tsubouchi2008critical,evans2010collaborations}.

\section{Industrial Experience and Tooling}
\label{sec:tools}

After a BSc and MSc in Computer Science, I did work for a few months as a software engineer dealing with database applications.
I then quit, and started a PhD on evolutionary computation applied to software engineering problems, followed by a post-doc on model-based testing.
That post-doc experience was of particular interest for the topic of this paper, as done in collaboration with two industrial partners.
What was really interesting is that I did quit that 3-year post-doc after 2 years to join one of those two companies as a regular software engineer. 
As I was knowledgeable on the details of that research project, I was dragged back into it
for its remaining final year.
However, this time as an ``industrial partner'', effectively experiencing both sides of the \emph{barricade} in the same project.

The following five years, after I quit the post-doc, I worked as an engineer and tester on quite a few different systems, like for example real-time ones controlling hundreds of thousands of embedded sensors, GUI applications dealing with complex 2D graphics, scientific computation and large scale web applications communicating with tens of web services.
Still, I always kept a foot in academia, being involved with some research projects (e.g., EvoSuite~\cite{fraser2011evosuite}) and having a part-time position as research fellow at the University of Luxembourg. 
Finally, I came back and accepted a position as associate professor in an university college in Oslo, Norway. 
During my career in industry, I mainly developed software in Java, with some parts of C++ (roughly one year), and a little bit of C\#, JavaScript and SQL.
I will therefore focus on my experience in Java, as for the other languages I consider myself just an amateur.

During those five years in industry, I have lost the count of how many thousands of test cases I had to manually write. 
It would had been helpful to use tools from academia that promise to generate tests automatically, or at least help at writing and managing them. 
Unfortunately, that was not really possible.  
However, there are plenty of tools (often open-source) that are ``industry-ready'' and can be used today by engineers to help writing and evaluating test cases.

In this section, I list what I actually used during my daily job. 
Looking at why those tools were developed, and by who they are maintained, provides some insight on what could be done in academia to achieve such kind of practical, engineering success.
As the lack of usable tools is one of the main barriers to knowledge transfer from academic results to industrial practices, looking at the existing tools is an important first step to address this issue.

A summary of the used tools is presented in Table~\ref{tools}. 
More details on these tools can be found in appendix.
All of these tools are open-source, although they might provide as well a pro version that requires to buy a license.
Many of these tools were started by single engineers in their spare time, and then grew in the open-source community.
As the lack of time is often cited as one reason for why researchers do not develop usable, engineered tools, it is important
to see how important the open-source community was for the success of those tools.   

\begin{table}
\centering
\caption{\label{tools} Summary of the discussed tools. They are all open-source. 
IntelliJ is the only one that also provides a pro, commercial version.
FindBugs is the only tool that started in academia.
For the cases in which the information was available, the table also specifies whether those tools were started by individual efforts of some engineers, instead of  teams of developers.}
\begin{tabular}{l ccc l} 
\toprule 
Name & Open Source & Academic & Individual & Short Description \\ 
\midrule 
 IntelliJ IDEA &  $\checkmark$ & &   & IDE\\
 JUnit &  $\checkmark$ & & $\checkmark$ & Test case framework\\
 Maven &  $\checkmark$ & & & Build tool\\
 Jenkins &  $\checkmark$ & & $\checkmark$ & Continuous integration\\
 JaCoCo &  $\checkmark$ & & $\checkmark$ & Code coverage\\
 Selenium &  $\checkmark$ & & & Browser tests\\
 WireMock &  $\checkmark$ & & $\checkmark$ & Web service mocks\\
 REST-Assured &  $\checkmark$ & & $\checkmark$ & REST API tests\\
 Mockito &  $\checkmark$ & & $\checkmark$ & Unit mocks\\
 ZAP &  $\checkmark$ & & & Penetration testing\\
 JMeter &  $\checkmark$ & & & Performance testing\\
 FindBugs &  $\checkmark$ &  $\checkmark$ & $\checkmark$ & Static analysis\\
\midrule
 Total   & 12  & 1 & 7 & \\ 
\bottomrule 
\end{tabular} 
\end{table}

Open-source development is a key for the success of tools used in practice.  
These tools are often started by single engineers in their spare time.
Development and testing can become easier when you can get contributions from
the open-source community.
But you can get help from engineers only if what you are addressing is of interest to them. 

\begin{result}
Academic tools should be released as open-source whenever possible.
\end{result}

\section{Example Problem Scenarios}
\label{sec:problems}

In the previous section, I have listed some of the tools I actually used during my five years in industry.
Those tools significantly helped in the development and testing of the software  systems I was working on.
However, during those five years I worked in industry as an engineer and tester, there were several cases in which I faced challenging testing problems. 
In many of them, automation would have helped significantly.
In others, there was simply no working solution. 

In industry we still face many concrete challenges when developing and testing software. 
I will discuss what I faced, and list these open problems, focusing on the three main ones.
This can be useful to show possible directions for future research on concrete, industrial problems that are still waiting for usable solutions.

Note that I worked in three different companies, whose names are not really important for this report. 
Due to confidentiality, even if they are former employers, when I will report on some anecdotal story I will not specify the company involved.

\subsection{Unit Test Generation}

Writing test cases takes a lot of time, and often it is not systematic, potentially missing important scenarios~\cite{Mye79}.
There are different kinds of testing, from concentrating on single units (e.g., Java classes) to address whole applications.
Unit testing is often one of the first steps when testing a developed system, and there 
is a large amount of work in the academic literature about how to automate it. 

However, among the many prototypes, there were only two tools that were mature enough to be actually used in practice:
the open-source academic Randoop~\cite{PLEB07} 
and 
the commercial AgitarOne\footnote{http://www.agitar.com}.  
Neither could be used in the different companies I worked in.
Randoop does generate JUnit tests, but, last time I tried, it does not provide any form of protection (e.g., a security manager) when dealing with the environment (e.g., if you have classes reading and deleting files, you might end up with all kinds of side effects when those classes are tested with random inputs).
AgitarOne generates tests are not meant to be ``readable'', and so it is unclear what one should do with those tests once generated (cannot be used for debugging for example). Furthermore, it requires way too complex setup.

For these reasons, I have been collaborating on the open-source EvoSuite~\cite{fraser2011evosuite} tool, which tries to generate unit tests automatically by using search algorithms (e.g., genetic algorithms). 
During these several years (EvoSuite started in 2010), a lot of improvements have been carried out.
I have been trying to use it on the systems I had at work, and that feedback was important to drive some of its development. 
For example, without the need to run it on actual systems in a production environment, likely we would not have developed plugins for Maven and IntelliJ~\cite{evo2016Maven}.

Did I, and my colleagues, use EvoSuite regularly at work in our daily job? 
The answer is \emph{no}, at least not yet.
Although thanks to the plugins for Maven and IntelliJ there is no major usability obstacle (unless of course you use Gradle and Eclipse),  there are other problems that need to be resolved first.
For example, a current showstopper is that some of the generated tests do wrongly fail at runtime.
Even if that happens only in 1\% of the generated tests, when you consider that on average EvoSuite generates roughly 14 tests per class, this leads to major issues.
If you run EvoSuite for the first time on a system having 1000 classes, then you end up with more than 100 useless failing tests, often in different files. 
Deleting those tests manually is a long a tedious task (I tried once on a 7000 class system, but then trying to remove manually all of the hundreds of failing tests was just a futile attempt).
As long as you have failing tests, you cannot really add them to the build in the version control system repository.

Note: such issue of failing tests is just a technical one. 
It just shows that there are still some bugs and edge cases that need to be taken care of, and fixing them is just a matter of time and resources. 
However, what is the incentive for academics in software engineering to fix this type of issues compared to publish more papers?
In the end, you can still publish papers even if you have problems with 1\% of the generated tests.
This is in clear contrast to the tools discussed in Section~\ref{sec:tools}, where the main driving force behind their development is to obtain useful engineering results for the practitioners.

There has been a lot of research in unit test generation, and it is an important topic for practitioners. 
However, what is really missing is not  novel, better techniques, but rather the engineering effort and incentives to make
such research techniques of widespread application in industry.

\begin{result}
The usability of academic tools should be an important factor to consider when developing them. 
\end{result}

\subsection{Environment Simulators}
\label{sec:simulators}

Unit testing is only one aspect of software testing. 
To increase our confidence on the correctness of the developed systems, we also need to carry out system testing,
whose details will depend on the type of system, e.g. embedded systems or web applications. 

When I was working as a post-doc in a model-based testing project, that work was done in collaboration with a company, where we used some of their systems for a case study for system-level test case generation.
At a high level, the idea was as follows~\cite{zohaibPhd}:
use UML to model the environment~\cite{iqbal2015environment} (e.g., sensors and actuators) of the system under test (SUT),
automatically derive executable Java code from such models,
and then use search-based testing to guide the simulation of environment events that lead the SUT in an erroneous state~\cite{AIB10}.

The developed prototype showed the feasibility of the approach.
However, due to several reasons, there was never a proper technology transfer, and the prototype was never actually used by the software engineers after the project ended.
As I worked in that company for the following three years, spending large part of my working time developing environment simulators for testing purposes, I had all of the motivation and knowledge to use such prototype if it worked.
If there was potential, I could have even managed to convince my manager to spend some of my working time in improving that prototype.

However, the main showstopper was that \emph{using UML was a far, much more time consuming and complex activity then just directly writing the simulators manually}. 
As soon as you start to have non-trivial models, and you put concurrency into the mix, understanding what is going on when you have problems becomes nearly impossible.
For example, if you automatically derive Java code from UML, and use such code to run the tests (e.g., in this case the environment simulator), then you want to use a debugger that works at the UML level.
Having to step into automatically generated code
that is hard to understand
would make debugging much more difficult,
likewise you would like a debugger that works at C/C++ level and not on compiled assembly code.
Furthermore, if asking for code completion is too much, at least you would like some static compilation checks directly in the IDE, and not having OCL constraints and Java snippets on the UML actions being just free text 
(that was in 2012, and current UML tools might have  improved by then).
Note: these issues are mainly technical and related to the usability of UML (other modeling technologies could had been better in that context), and they have been discussed at length before~\cite{whittle2013industrial}. 
Furthermore, although these issues might be a showstopper for a beginner, they might be a lesser issue for an experienced UML user.
Still, if those problems are not addressed, technology transfer from academic research to industrial practice would not be possible.

At any rate, there were few cases in which having models (not necessarily UML) would had been highly desirable.  
Once, I was moved to a new project  where hardware components were involved, and I had to learn how such hardware worked before I could start working on their software.
There was no documentation.
The only person with knowledge of the system was the previous engineer that worked on that system.
The knowledge transfer was made during a meeting where a white-board was used.
To explain the system and its interactions with the hardware components, a finite state machine was drawn on the white-board.
It would had been much more effective to formalize such state machine in a model, and then have tools to derive code and test cases  automatically from it. 


\begin{result}
Using environment simulators to effectively test systems of systems is a very important topic in industry. 
\end{result}

\subsection{Testing of Web Applications}

In my last assignment before going back to academia full time, I worked for more than a year as a test engineer for a web application.
Most of my working time was spent in writing system level test cases in Selenium, and at times with REST-Assured when testing some web services.
However, when dealing with this type of systems, the real complexity lies in setting up their environment.
This means using embedded databases which are initialized from the tests by executing SQL commands, and also having to configure mock responses in WireMock for every single call to an external web service.
Even on a simple Selenium test case doing at most 4-5 mouse clicks in the browser, you would end up spending the next two working days configuring the required 20-30 web service connections, each one using a large amount of XML/JSON data. 

How great would had it been to generate those tests automatically?
Unfortunately, there was no tool/technique that I could use.
There are some tools that can create sequences of events on a GUI, but those would be useless in this context. 
This a very important engineering topic that requires more investigation.

Besides controlling an environment (e.g., databases and web services) directly in the tests, another complementary approach is to use a real environment, in a kind of hardware-in-the-loop testing. However, in this context one cannot really know what is already present in the databases, and how the web services will behave, as that might change without warning (especially the data in the databases).
This is the case in large organizations where a team working on a system has no control over the other systems, and a test infrastructure for ``live'' integration testing of all systems might change often, e.g. at each new component release. 
Still, in such cases one can  use some sort of automated test sequence generation on the GUI, although there would be the issue of what to use as ``oracle'': in a web application running on a server like Tomcat or Jetty, it is very unlikely that a user input would crash the whole application.
However, one thing that can be done is to look at the application's logs, where one would not expect to see any ``ERROR'' message, i.e. they can be used as oracle in this context.

To do this type of testing, one could use a web crawler that works in the browser, and then check the logs after the crawler is terminated.
To do this, I did try to use Crawljax~\cite{mesbah2008crawling}, a popular academic web crawler.
Unfortunately, I did not manage to make it work on the system I was 
testing.\footnote{https://github.com/crawljax/crawljax/issues/496}
However, writing a random testing~\cite{AIB11} tool was quite simple, and quite effective at the same time.

Another big issue I faced when dealing with testing of web applications was how to measure the effectiveness of  Selenium tests.
One could of course calculate the code coverage of the server side code.
Interestingly enough, this was not really possible in Java until version 0.7.7 of JaCoCo came out in June 2016 (all Java code coverage tools were only for unit tests, and could not be used for system tests in a multi-module project). 
One could use the commercial Clover, but, besides being quite expensive, it has its own set of limitations, as it does instrumentation at source code level and it does change the structure of the instrumented classes (this is a big issue when having code relying on reflection, like JSON parsing libraries). 
Another complementary approach is to formalize the requirements of the system, and then map each requirement to one or more system level test case (and so check if some requirements are not covered by any test).
Even if those two complementary approaches work well, still remains the fact that the actual  testing of links/forms in the HTML pages is not really measured, and neither the JavaScript code in them.
This is of particular importance in web applications where parts of the HTML structures are dynamically generated on the client side using JavaScript.
Knowing the quality of the current test suites is very important for managers when deciding on resource allocation for validation and verification tasks.

To get a better picture of what the Selenium tests were covering, I investigated if there was any tool that gives some information on what gets covered in the HTML pages.
The only tool I found was the academic DomCovery~\cite{mirzaaghaei2014dom}.
Unfortunately, currently it has not been updated since 2014, and the current version on GitHub does not compile due to a snapshot dependency that does not exist anymore.
This is definitively a very important topic that warrants more research and engineering effort on.

\begin{result}
Not only the automated generation of tests for web applications is lacking, but also the tools and metrics to quantify the
effectiveness of the existing manual tests are lacking as well.  
\end{result}

\section{Examples of Popular Topics in Academia}
\label{sec:popular}

In the previous section, I have discussed some open problems in software testing that I faced while working in industry,
and for which there was no mature solution I could use. 
More research on such topics, with usable engineering solutions, would be very important for practitioners.  
However, on what different topics do academics rather prefer to work on?

While working in industry, I have been still involved with academics, either with direct collaborations, or as reviewer for conferences and journals. 
Furthermore, still once/twice a year I was attending academic conferences. 
There are a few topics that are quite popular in academia, with hundreds if not thousands of scientific articles about them. 
However, in my daily work I have never dealt with them, although in theory the type of systems I have been working on would had been natural use cases for them. 

I will here discuss some of these topics, and why they did not apply to the systems I have been working on, providing some thoughts on why that was the case.
This feedback should be useful for researchers when trying to understand industry contexts and requirements. 
Such selection of topics is based on my personal experience when dealing with academia (e.g., conference attendance and review requests).
However, such selection of topics is in line with recent analyses on what topics are mostly discussed in academic conferences
in contrast to the practitioner ones~\cite{garousi2017}.

Note, however, that the fact I did not deal with those topics does not mean they are not important, or that they would not be useful in other contexts or companies.
There are many, many companies out there, each one with its own set of constraints and needs.
For example, the needs for developing an embedded system are likely quite different from the ones for developing a web application.
The goal of this section is to provide more insight on these topics, to better understand their strengths and limitations.

\subsection{Regression Testing}

There is a large body of work on regression testing~\cite{yoo2012regression}, in particular on \emph{selection} and \emph{prioritization}.
I never had the need to deal directly with any of those.
The reason is simple: either I dealt with systems with no or very few automated tests, or the tests were fully automated.
This latter case is quite important: let us consider a test suite that takes 8-9 hours to run (which was the case in one of the projects I worked in).
As a developer, if you make a change, you do not really want to wait 9 hours to find out that you broke something. 
At a first look, regression testing prioritization could sound useful here.
Well, that is the case until you realize that nowadays hardware is very cheap compared to employee salaries, and having a continuous integration server running 16 tests in parallel does cost only very little in comparison.

It could be argued that even if you reduce execution time from 9 hours to half an hour, then 30 minutes is still not a negligible amount of time.
That is the case until you start to take into account the daily routines of software developers.
It is not uncommon that, when you change one part of the code, you run all the tests involved in that functionality you are modifying. 
And that usually does not take half an hour. 
Furthermore, before pushing a change (e.g., in Git or Mercurial), you would still run all the unit tests anyway as part of the build. 
There is still of course the possibility that you might break some unrelated functionality. 
But, even in that case, it might not be a big issue.
The continuous integration server will tell you which tests are now failing, and then you can try to fix the regression bugs and run just those failing tests during debugging. 
Worst case, considering revision control system like Git, you can just revert your changes, unless you were already working on a private branch.
When you consider a 7.5 hour day work, where there are anyway breaks for meetings, coffee, lunch, etc., a half an hour build is not such a big issue, especially when you can do other tasks meanwhile.
Of course, if you need to release new software versions to clients every day, then 30 minutes is a problem.
But, in many cases (e.g., in all the companies I worked at), release cycles are counted in weeks/months.

There are, however, cases in which regression testing optimization would be highly desirable. 
For example, if you have millions of test cases, then regression test prioritization would be essential (although that does not seem a common case).
Furthermore, if you are doing hardware-in-the-loop testing, and the involved hardware is very expensive, then parallel test execution might not be a viable option.
In some other cases, you might not even have automated tests.
In one project I worked with, for example, before doing a quarterly release, the software had to go through a QA process, where it was tested against actual hardware.
Being a very large and complex system of systems dealing with a lot of different hardware components, there was no automated test for the whole system. 
The manual testing process had to be executed by two domain experts, taking usually between two and three \emph{weeks}, following a list of test case instructions written in a Word document.
If there is a critical bug, you really do not want to find out after three weeks, as that would delay the release to the clients. 
Unfortunately, I could not recommend them any regression testing technique that would have helped in such a context, as there is none that I know that could had been used.

When doing experiments in regression testing, it is  very important to take in consideration how long does the test suite take to run, and if it is fully automated or not (i.e., can its execution be easily parallelised?). 
If it takes only few minutes, or even worse just few seconds (e.g., if the tests are just unit tests), then the representativeness and relevance of such experiments could be put into question.  

\begin{result}
Regression testing optimisation might be not so important when test suites are cheap and fast to run.
\end{result}

\subsection{Model-Based Testing}

I did a post-doc in model-based testing (MBT) for more than two years~\cite{AIB10,IAB10,iqbal2015environment}. 
And there is large literature on modeling, especially when considering the more general context of model-driven engineering (MDE)~\cite{schmidt2006model}. 
In a recent analysis on what topics are covered in academic conferences, \emph{model} was the most common word in the talk titles~\cite{garousi2017}. 

However, the two companies involved in that research project had no models: it was me and the student in the project that did all the modeling.
None of the companies I worked for in the following five years used any sort of modeling.
When looking for new jobs, and searching for positions that required knowledge in Java and testing (my main areas of expertise), I have never seen a job post listing UML as a requirement or desired skill (of course, if you search for UML, you do find some job descriptions asking for it).
When doing interviews in different companies (spanning from seabed exploration to fitness equipments and streaming music  services), no interviewer ever asked me about UML or modeling in general.
And I have never encountered an open-source project that uses UML.
This is not surprising, as it is estimated that just 0.28\% of open-source projects uses some UML~\cite{hebig2016quest}, where 2/3 of them only contain a single UML file. 
Furthermore, based on a survey of 3785 developers, design models seem rarely used in practice~\cite{gorschek2014use}:
\begin{quote}
``The use of models in general, and the UML in particular, does not seem to be standard practice and de facto standard in software development,  which challenges the assumption on which much of current research is based.''
\end{quote}


However, there is no doubt that modeling is important: one can just look at 
Simulink\footnote{http://se.mathworks.com/products/simulink} 
for example, and see how it is the de-facto standard in many automotive domains.
Also 
IBM Rational Rhapsody\footnote{http://www-03.ibm.com/software/products/en/ratirhapfami} 
seems being used in quite a few embedded-system domains. 
And if there are companies that make a business out of selling UML tool licenses, then it means that there are people that find it useful enough to pay for it.
Furthermore, a few times I wished I had a working modeling solution (recall Section~\ref{sec:simulators}).  
But I always wondered how much widespread and useful is MBT on UML models, and MDE in general, at least for the types of \emph{enterprise}, non-embedded systems I have been working with.
If it is really so useful as supporters say, one would expect a much wider adoption, especially considering how many years have passed since UML was introduced in the 90s.


When doing research on UML/MBT, and doing empirical studies to validate new theories and techniques, it is hence important to always state where the models come from: are they artificial or did they already exist before the experiments?
In the former case, one has to argue and evaluate if the time and effort in developing and maintaining such models for a given system does pay off in the end~\cite{boehm1981software}.
Reporting on success stories in industry would be very useful to cast out the doubts from skeptical readers. 
If UML/MBT is of wide applicability, and if it does bring a lot of benefits in the long run, there should be more emphasis in publicizing it with \emph{concrete} success stories, and make engineers know about it, which would lead to a larger adoption.
For example, one way forward would be by spreading the word at developer conferences and by engaging with the open-source community, besides making the MDE tools more user friendly~\cite{whittle2013industrial}.
Otherwise, one would end up agreeing with what magazines like the American Scientist state about UML~\cite{as11}:
\begin{quote}
``$\ldots$ in the 1990s a group of respected software designers combined forces to create a graphical notation for computer programs called the Unified Modeling Language (UML), which was intended to fill the role of blueprints and circuit diagrams in civil and electrical engineering. Despite a great deal of hype, UML never really caught on: Almost everyone who earns a degree in computer science learns about UML at some point, but very few programmers use it voluntarily $\ldots$''
\end{quote}

\begin{result}
The use in industry of UML does not seem to be as widespread as large part of the academic community seems to believe. 
\end{result}

\subsection{Mutation Testing}

You can write test cases without a single assertion. 
The test cases would still achieve the same degree of code coverage,
but would only fail if an unexpected exception is thrown.
For regression testing purposes, this kind of tests would have 
limited effectiveness.
To evaluate how good the assertions in the tests are, one can use
what is called \emph{mutation testing} (MT)~\cite{mutation2011}:
the idea is to inject faults in the system under test, and then
see if the test suite is able to catch them (i.e., tests that
pass on the correct version should now fail on the buggy one).

However, one of the main limitations of MT is the so called 
\emph{equivalent mutant} problem: an injected fault might
not result in any actual fault, as it could be just a syntactic
change that does not alter the semantic of the code.
Unfortunately, detecting equivalent mutants is an undecidable problem.

During my work in industry, I have never used MT, although
there are  tools like 
PIT\footnote{http://pitest.org} that I could have tried.
Point is, I always worked on systems in which there was a clear need to have more tests,
because even the results for basic coverage criteria like statement coverage were not satisfactory.
When large parts of a system are not covered with automated tests 
(although still tested through manual QA), 
you do not really need more sophisticated adequacy criteria like MT.
One can still add PIT in Jenkins (configuring it takes just a few minutes),
and get reports from it.
But, due to the equivalent mutant problem, the absolute values of the mutation scores   
are not so useful. 
Still, PIT is used by some practitioners (currently nearly 400 stars on GitHub).
One hypothesis is that, when developing complex code, and an engineer might want to be very
sure that the code is correct, s/he might want to spend extra time in manually checking  things like the missed / non-killed mutants.
In other words, the context of using MT during software development can be different from the context of continuous integration and regression testing.

The case of PIT shows a wonderful example of collaboration with industry and academia~\cite{pit2016}.
From the information available online (e.g., Github), it looks like PIT has been mainly developed by an engineer in industry, likely influenced by the work done in academia (there are literally hundreds of scientific articles on MT). 
And PIT is not a throw-away prototype. 
It is an actual tool with a lot of engineering effort behind it, shown for example by its integration with Maven, Gradle, Ant, Eclipse, IntelliJ, etc.
One can assume that the  co-authoring academics~\cite{pit2016} will help to make the tool even better, by delivering the results of cutting edge research through a tool actually used by many practitioners in industry. 
Such a success story should be of inspiration for any researcher working in software engineering. 

However, MT is most useful when you have high quality/coverage test cases, and then you want to have guidance on what more you should add.
As the lack of such kind of tests is a problem in many enterprises, the importance of MT might currently be not as high as it could otherwise be.
With better automated test case generation in the future, MT could become more important, especially considering that mutation testing can be used to drive the automated generation of tests~\cite{mutation2015emse}.
Given a set of automatically generated test cases that achieve high statement coverage, MT could be used to decide which further tests should be added manually.

\begin{result}
Albeit important, after many decades Mutation Testing is still not widespread in industry.
\end{result}


\section{Discussion}
\label{sec:discussion}

\subsection{Impact Metrics}
As discussed in Section~\ref{sec:rw}, there are already some academics that realize the lack of impact of academic research on practice, which is a very bizarre situation for an \emph{engineering} field.
When moving to industry after  a PhD/post-doc, there was not much that I learned from research that I could directly use in my daily job.
However, there are plenty of available techniques and tools that are developed in the open-source community which I did use (Section~\ref{sec:tools}).
There are plenty of open problems in industry that still need to be solved, and that could benefit from academic involvement (Section~\ref{sec:problems}).
In particular, apart from random testing, I could not use any form of automated test data generation in my work, which could have saved a significant amount of effort and resources.
And automated test data generation is a topic that has been addressed in the literature for decades.

There is not an easy solution about the lack of impact of academic research, otherwise it would had been found already long ago.
As stated before by few academics, reward systems that take into account the \emph{impact} on practice should be introduced. 
If researchers in software engineering cannot change how universities are run (e.g., promotions and tenure systems, especially when software engineering is part of the mathematic or computer science department), at least they do have direct control on the software engineering conferences and journals.
If putting impact on practice as a publication requirement would be too strong, at least impact should be a major factor when considering incentives like awards and prestige.
For example, when deciding for a best paper award among a selection of papers, whether they provide downloadable tools that can be evaluated by the committee should play a major role in such decisions.

Furthermore, in many software engineering venues there are separated tracks for tools and applications in practice, which often are treated as second-class citizens  compared to the ``main'' research track.
At times, they do not even have awards, or, even worse, I have seen conferences where the presentations for tool papers/demonstrations were canceled or squeezed to make more space for the research presentations. 
Those tracks should have the same importance of the main research track, if not more.  

In an engineering field, it could be argued that papers presenting actual results of engineering research on practice  are more fitting than blue-sky ideas that might take 15-20 years before reaching engineering maturity (if ever at all).
And, when considering new techniques that are empirical evaluated only on toy-problems, those should not be put at the same level of actual engineering results in industrial contexts, even if these latter are on problems considered ``solved''  by academics (which often translates to the existence of previously published papers empirically evaluated on just some small examples, where the \emph{scalability} to real systems was left as ``future work'' that never followed). 
Another option is that, if this is not possible or even wanted, at least those conferences/journals should remove the word \emph{engineering} from their name, or explicitly add the word \emph{research}, as to avoid confusing the engineers that might want to attend those conferences or read those journals.

Funding agencies also plays a major role in the lack of research impact on industry:
``If funding projects changed their metrics and allowed more money and time to be dedicated at developing proper, usable, tools I think that many researchers would do so. In the current environment the money/time is only enough for a prototype at best''. \footnote{Anonymous reviewer.}
As usable tools would be of direct benefit for the taxpayer, especially when compared to research articles behind paywalls, 
funding agencies could be convinced to better fund such type of endeavors.

\subsection{Open-Source Contributions}

All tools and techniques that I used in my daily job as an engineer and tester were developed by engineers as open-source  (Section~\ref{sec:tools}).
Therefore, I strongly believe that, to improve achieving impact, academics should get more involved with the open-source community.
This is not just a matter of releasing the prototypes as open-source (which would also help in making research results repeatable~\cite{collberg2016repeatability}), 
but also to get into contact and collaborate with engineers that do spend part of their free time working on open-source projects.
This is particularly important, especially when considering that the steps from prototype to tool that are actually scalable and usable do require a non-negligible amount of engineering work and knowledge. 
If you look at the millions of projects on open-source repositories like 
Github\footnote{https://github.com}, often developed by software engineers in their free time, there is plenty of potential there, as long as you work on topics that software engineers consider useful~\cite{lo2015practitioners}. 
Research contributions to existing software projects used by hundreds of thousands of people, if not even millions in some cases,  could have more impact than writing yet another dozen of research/vision papers on blue sky ideas that maybe only just a handful of PhD students will ever read.
As contributions to open-source projects are easily trackable (e.g., all commits from each author are visible), and the popularity of projects can be somehow measured (e.g., in number of stars in Github, where the over 37,000 stars of
Linux\footnote{https://github.com/torvalds/linux} 
can be taken as a reference point),
such contributions should be considered among the ``applied research metrics''.

Of course, the development of open-source tools should  not be the only main measure for assessing impact. 
There is a large of body of research on empirical studies that would not result in building any tool, albeit such work can be very important when addressing real problems in industry.
For example, a very common topic in industry is \emph{Test Driven Development}, where empirical studies made in academia can help to better understand its benefits and downsides~\cite{rafique2013effects}. 
Unfortunately, measuring the impact of such work in an objective (even if partial) way is hard.

\section{Threats to Validity}
\label{sec:threats}

What reported in this paper is based on my personal experience of working five years
in different companies.
As such, what reported cannot of course be generalized to all kinds of companies and testing contexts.
To a certain extent, what reported in this paper could be arguably considered just as a set of anecdotal stories and personal opinions, as based on the industrial experience of only a single engineer and his interactions with his colleagues in three different companies.   
To get a better, more precise understanding, many more companies and engineers should be involved to study what they use in practice and how academic research affects them.  
However, as such type of experience report from engineers in industry is not common in the literature, more is needed to collect over time a large enough body of knowledge from which we can draw reliable conclusions.

When speaking about tools, it might well be that I missed some, or that my negative experience with some of them was just due to my lack of understanding and inability to properly use them.  
In this latter case, however, this might point to put better emphasis on documentation and usability concerns.
Furthermore, as I have been mainly working on systems written in Java, I mainly focuses on tools related to Java. 
The context in other languages (e.g., C\# and Python) might well be very different.

At times, it might be difficult to give proper credit where it is due. 
If some practitioner tool was influenced by academic research, then it is not always so simple to find out about it. 
Consider for the example the case of PIT: an engineer using it might have no clue of the large amount of research done in academia on mutation testing.

\section{Conclusion}
\label{sec:conclusion}

Arguably, research in software engineering, and particularly software testing, has had only limited impact on current practices in industry.
Ideas born in academia can influence current practice after a couple of decades, but usually the engineering challenges of implementing such ideas and make them \emph{scalable} for real-world software engineering problems is often done in industry.  
For an engineering field, this should not really be the case.

Such topic has been discussed at length in the literature, where the possible causes have been discussed several times.
One major problem is that, often, software engineering is wrongly treated as a scientific field, and not as an engineering one.
Furthermore, academics are rewarded based on metrics that mainly consider paper publications, and not engineering impact on practice.

In this experience report, I have shared my experience of working as an engineer in industry for five years after a PhD and a post-doc in software testing.
What could I use in my daily engineering work from what I learned in my previous five years in academia?
Turned out, very little.
Again, for an engineering field like software engineering, this should not be the case.
There are many open problems in industry, that for example me and my colleagues have been facing when developing systems in several different domains.
Testing is a major one, where automatic test generation would be a very desirable possibility/feature.
Unfortunately, apart from random testing, there was nothing else I was able to concretely use in my daily work.

There is no easy solution for the limited impact of software engineering research on practice. 
Putting more emphasis on the engineering side of software engineering is a first step in the right direction.
Closer collaborations with industry would help at focusing on important topics of practical value. 
Getting more involved in open-source projects that can reach millions of people would be a natural following step.
In any case, it is essential to remember that software engineering and computer science are not the same thing: both are important, but the expectations should be different.

\section*{Acknowledgments}
I wish to thank Lionel Briand for insightful discussions.
I also wish to thank
Per Lauv\aa{}s and Zohaib Iqbal
for useful feedback on an early draft
of this paper.
This work is supported by the National Research Fund, Luxembourg (FNR/P10/03).

\appendix
\section{Appendix}

\subsection{IntelliJ IDEA}

Like for production code, to write tests you need an editor.
One of the most popular IDEs for Java is 
IntelliJ IDEA\footnote{www.jetbrains.com/idea/specials/idea/idea.html} .
Simply put, the autocomplete feature in IntelliJ IDEA is the most important advancement in software development I have experienced since I started to write code in 2000 when I was a student. 
No longer continuous copy\&paste of long variable/method names, or even worse typing them  directly. 
Most of the time, just the first one/two letters are enough to correctly autocomplete. 
Plus there are all the other kinds of smart autocomplete, like for example autoclosing of XML tags, or showing an autofilled selection of valid data based on the current context (especially useful when dealing with Maven pom files).
Other IDEs have some basic support for autocomplete, but they look very basic in comparison. 

From a scientific point of view, arguably there is not much ``innovation'' in an autocomplete feature, as it is a relatively simple idea.
However, likely there are  many engineering challenges to make it right and user-friendly
in a general way applicable in different contexts.
Would  these engineering challenges be of interest for an academic conference? 
However, IntelliJ IDEA also provides cutting-edge techniques related to 
code-refactoring, and code-refactoring has deep roots in software
engineering research~\cite{refactoring2015}.

Academic research can have large influence on practice (e.g., refactoring), but there are often important topics, with direct impact on software development practice, that have received less attention (e.g., autocomplete). 
What is important for practitioners does not always match what researchers find interesting to work on.


IntelliJ IDEA is developed by a Czech company called JetBrains, founded by three Java developers in 2000. 
The community edition is open-source, although the pro version (needed for enterprise programming) requires to buy a license.

\subsection{JUnit}

To write and run tests, you need a framework.
In the Java world, the de-facto standard is
JUnit.\footnote{http://junit.org}
There are other popular frameworks as well, like TestNG and Spock, but none of the projects I worked in used them. 
JUnit was founded by a software engineer and a computer scientist (currently working at Facebook and Microsoft).

\subsection{Maven}

To compile a Java project and handle all the third-party libraries automatically, you need a build tool.
Currently, the de-factor standard is 
Maven.\footnote{http://maven.apache.org}
Other options are Gradle and Ant.

Of particular interest is the fact that build tools are used to run test cases as well,
e.g. by executing the command ``mvn test''.
Maven has two main official plugins dealing with the running of tests: Surefire (for unit testing) and  Failsafe (for integration testing).
These plugins provide a lot of extra functionalities, like for example the ability of running tests in parallel or re-executing flaky tests up to a certain number of times.

Maven is developed by the 
Apache Software Foundation\footnote{http://www.apache.org}, 
a non-profit corporation with apparently no employee but thousands of volunteers.

\subsection{Jenkins}

When several developers work on the same piece of software, it is important to have a remote Continuous Integration server, where the system is built at each new pushed code change.
Each time the system is built, all the regression tests are automatically run as well.
This is also important for long to execute test suites, which might be cumbersome otherwise to run directly on the developers' machines.

Arguably,
Jenkins\footnote{https://jenkins.io}
is the most popular Continuous Integration server for Java. 
Another one is for example Bamboo.
Jenkins has several utilities and plugins to help monitor the execution and collect the results of the test cases. 
  
Jenkins is an open-source fork of Hudson, which was originally developed at Sun Microsystems (the company behind Java before being acquired by Oracle).
The lead developer of Jenkins/Hudson is an engineer now working as CTO at CloudBees, 
a company that provides  continuous delivery solutions powered by Jenkins.

\subsection{JaCoCo}

How many tests one should write? 
Is the current test suite ``good enough'' or should it be extended?
Of course, one  can (and should) have a test for each requirement, but that
does not tell you what in the code has been tested. 
A complementary approach is to look at what is executed by the tests, and
then check if any part of the source code has not been exercised. 
If some parts of the code is not executed, it would be good to add tests
to cover them, as those parts might have bugs.

One of the most used tools for Java code coverage is the open-source
JaCoCo.\footnote{http://www.eclemma.org/jacoco}
Others are Cobertura and the commercial Clover.
JaCoCo was started by an engineer now working as CTO in a consulting company.

\subsection{Selenium}

Writing system level tests for a web application is a challenge, as users
interact with it by using a browser (e.g., Chrome or Firefox).
The 
Selenium\footnote{http://www.seleniumhq.org}
tool simplifies this task by making possible to programmatically control browsers
directly from the test cases.
Selenium was originally started at ThoughtWorks as an internal tool to simplify their testing activities.

\subsection{WireMock}

Many enterprises today rely on web services (e.g., SOAP and REST), especially 
when using microservice architectures.
Writing deterministic tests against applications using web services is challenging,
as usually the developers of such applications do not have control over those
external web services.

WireMock\footnote{http://wiremock.org}
 is a tool that starts a proxy server, where HTTP responses can be easily
configured based on the incoming messages. 
This can be used to practically mock away those external web services without the need
to modify the tested application.
WireMock  is developed and maintained by an engineer currently working as a consultant. 

\subsection{REST-Assured}

When testing a RESTful web service, one has to craft an HTTP message, open a TCP
connection, send the message, read back the answer, and finally evaluate assertions 
on such response.
To greatly simplify this sequence of operations, one can use a library like
REST-Assured.\footnote{http://rest-assured.io}
REST-Assured  was started by an engineer working as a consultant.

\subsection{Mockito}

When unit testing code, there is often the problem of dependencies that need complex initialization.
At times, some classes cannot be even instantiated, as only interfaces are available.
A solution for this kind of problems is to create mock objects for those dependencies.

In Java, there are several libraries that can create mocks, like EasyMock and JMock.
The most used one is 
Mockito.\footnote{http://mockito.org}
Mockito  was started by an engineer working at ThoughtWorks.
As of 2013, Mockito is among the top 10 libraries used in Java among open-source projects.
\footnote{http://blog.takipi.com/we-analyzed-30000-github-projects-here-are-the-top-100-libraries-in-java-js-and-ruby}

\subsection{ZAP}

Security is an extremely important topic when dealing with web applications.
It is not rare that security breaches end up in the newspapers,  e.g. when hackers steal large number of personal details, like the 500 \emph{million} user accounts stolen at Yahoo\footnote{https://www.theguardian.com/technology/2016/sep/22/yahoo-hack-data-state-sponsored} in 2014.
Often, you just need a silly mistake, like forgetting to secure a single cookie,
to compromise a user account or a whole web application.

To help mitigating this type of issues, one can use the
ZAP\footnote{https://www.owasp.org/index.php/ZAP}  
tool, which can automatically do different kinds of security attacks against web applications.
ZAP is one of the main tools of the OWASP Foundation, a non-profit organization dedicated to web application security.
ZAP was started by an engineer now working as a security expert at Mozilla.

\subsection{JMeter}

Besides validating the requirements of a web application, there are also non-functional properties that need to be taken into account, like for example response time.
Would the application behave properly when under heavy load, like for example hundreds
of thousands of users accessing it at the same time?

To help writing performance tests, one can use
JMeter.\footnote{http://jmeter.apache.org}
JMeter is  developed by the members of the Apache Foundation.

\subsection{FindBugs}

It is not uncommon that developers make the same typical mistakes, like opening a resource and then forgetting to close it. 
Or calling methods on immutable objects like strings and then forgetting to save the result in a variable.
It is possible to define patterns of common mistakes, and then statically scan source code 
for those patterns without the need to execute the analyzed code.
One great benefit of statical analysis is that it can be easily applied on large code bases and find real errors relatively quickly.

The most famous static analysis tool for Java is 
FindBugs.\footnote{http://findbugs.sourceforge.net}
Another popular tool is SonarQube, which also provides a commercial version.
FindBugs was originally developed by academics at 
the University of Maryland~\cite{HoP04}.

\bibliographystyle{spmpsci}      

\end{document}